\documentclass[aps,prd,amssymb,twocolumn,superscriptaddress,floatfix,nofootinbib,10pt]{revtex4-1}
\usepackage{latexsym}
\usepackage[normalem]{ulem}
\usepackage[utf8]{inputenc}
\usepackage{cancel}
\usepackage{graphicx}
\usepackage{color}
\usepackage{natbib}
\usepackage{float}
\usepackage{amsmath} 
\definecolor{BurntOrange}{rgb}{0.8, 0.33, 0.0}
\usepackage{hyperref}

\newcommand{\ba}{\begin{eqnarray}}
\newcommand{\ea}{\end{eqnarray}}

\newcommand{\be}{\begin{equation}}
\newcommand{\ee}{\end{equation}}

\usepackage[x11names,dvipsnames]{xcolor}

\usepackage{slashed}
\usepackage{pifont}
\usepackage{dcolumn}% Align table columns on decimal point
\usepackage{graphics}                                                                         
\usepackage{multirow}
\usepackage{makecell}
\usepackage{bm}

\begin{document}
                                                                  
%\date{\today}
\title{Kaon-deuteron femtoscopy from unitarized chiral interactions}

%\author{ \`Angels Ramos, Juan M. Torres-Rincon$^1$, Alejandro de Fagoaga$^2$ }

% \affiliation{$^1$Departament de F\'isica Qu\`antica i Astrof\'isica, Facultat de F\'isica,  Universitat de Barcelona, Mart\'i i Franqu\`es 1, 08028 Barcelona, Spain and Institut de Ci\`encies del Cosmos (ICCUB), Facultat de F\'isica,  Universitat de Barcelona, Mart\'i i Franqu\`es 1, 08028 Barcelona, Spain}
% \affiliation{$^2$ Facultat de F\'isica,  Universitat de Barcelona, Mart\'i i Franqu\`es 1, 08028 Barcelona, Spain }

\author{\`Angels Ramos}\affiliation{Departament de F\'\i sica Qu\`antica i Astrof\'\i sica, Universitat de Barcelona, Mart\'{\i}  i Franqu\`es 1, 08028 Barcelona, Spain}
\affiliation{Institut de Ci\`encies del Cosmos, Universitat de Barcelona, Mart\'{\i}   i Franqu\`es 1, 08028 Barcelona, Spain}
\author{Juan M. Torres-Rincon}
\affiliation{Departament de F\'\i sica Qu\`antica i Astrof\'\i sica, Universitat de Barcelona, Mart\'{\i}   i Franqu\`es 1, 08028 Barcelona, Spain}
\affiliation{Institut de Ci\`encies del Cosmos, Universitat de Barcelona, Mart\'{\i}   i Franqu\`es 1, 08028 Barcelona, Spain}
\author{Alejandro de Fagoaga}\affiliation{Facultat de F\'\i sica, Universitat de Barcelona, Mart\'{\i}   i Franqu\`es 1, 08028 Barcelona, Spain}
\author{Esteve Cabr\'e}\affiliation{Facultat de F\'\i sica, Universitat de Barcelona, Mart\'{\i}   i Franqu\`es 1, 08028 Barcelona, Spain}

\date{\today}

\begin{abstract}
We have performed a theoretical study of the correlation functions of $K^- d$ and $K^+ d$ pairs and compared them with those provided by the ALICE Collaboration from Pb-Pb collisions, and also from high-multiplicity p-p collisions in the case of $K^+ d$.
In addition to implementing the effect of the Coulomb force, 
the $K^- d$ and $K^+ d$ wave functions are derived from the corresponding strong scattering amplitudes that are built employing a unitarized chiral model for the elementary $K^- N$ and $K^+ N$ interactions. 
We present results for the impulse approximation, which accounts for single-scattering processes of the kaon with the nucleons of the deuteron, as well as for the solution of the Faddeev equations in the so-called fixed center approximation, which includes multiple rescattering effects.
The $K^- d$ correlation function is shown to be very sensitive to both the size of the source and the relative momentum of the interacting pair, with large deviations from the Coulomb baseline and sizable multi-step scattering contributions, effects that are tied to a ${\bar K}N$ strong interaction that is dominated by the influence of the subthreshold resonance $\Lambda(1405)$. In contrast, the $K^+ d$ correlation function only differs appreciably from the Coulomb one for relatively small sources, reflecting the mildly repulsive and elastic behavior of the $KN$ strong force.
The calculated correlation functions are found to nicely reproduce the experimental data of the ALICE collaboration. Our study serves to reinforce the validity of the theoretical models employed and demonstrates the value of femtoscopy as a powerful tool for probing hadronic interactions involving strangeness.
\end{abstract}

\maketitle

%\tableofcontents

\section{Introduction}
\label{sec:intro}

Among the wide variety of hadrons produced in relativistic heavy-ion collisions (RHICs), light nuclei are particularly notable because of their composite structure and small binding energies~\cite{Braun-Munzinger:2018hat,Oliinychenko:2020ply}. At high collision energies, their yields are expected to be strongly suppressed according to the statistical thermal model~\cite{Andronic:2005yp,Andronic:2010qu,Andronic:2017pug}. Nevertheless, it has been possible to measure their production and quantify their spectra, as demonstrated by studies performed by the STAR collaboration~\cite{STAR:2009kaf,STAR:2021ozh,STAR:2022hbp} and the ALICE collaboration~\cite{ALICE:2015wav,ALICE:2017nuf,ALICE:2019bnp,ALICE:2021mfm} (see also~\cite{ALICE:2022veq} and references therein). 

The deuteron, with a binding energy of 2.2 MeV, is a paradigmatic example of a loosely bound state, with its yield measured in various RHIC experiments, e.g.~\cite{STAR:2019sjh,ALICE:2017xrp,PHENIX:2004vqi,ALICE:2019dgz}. The survival of this state in extreme conditions despite its small binding energy has been addressed by coalescence models~\cite{Csernai:1986qf,Scheibl:1998tk}. These invoke mechanisms such as detailed balance~\cite{Oh:2009gx,Zhu:2015voa,Sombun:2018yqh,Liu:2019nii,Oliinychenko:2020znl,Zhao:2020irc,Staudenmaier:2021lrg,Coci:2023daq}, clustering~\cite{Kireyeu:2022qmv} and collisional broadening~\cite{Neidig:2021bal}. In contrast, the statistical thermal model assumes that deuterons are formed at chemical freeze-out, with abundances determined by thermal weights~\cite{Andronic:2005yp,Andronic:2010qu,Andronic:2017pug}. The two frameworks represent fundamentally different pictures of light nuclei formation, and ongoing studies continue to explore their validity.

In particular, experimental tools are essential to distinguish between these two scenarios. Femtoscopy is a technique used to probe the space-time structure of the particle-emitting source, as well as the interactions between hadrons~\cite{Koonin:1977fh,Bauer:1992ffu,Baym:1997ce,Wiedemann:1999qn,Heinz:1999rw,Lisa:2005dd,Fabbietti:2020bfg}. In the context of deuterons,  femtoscopy has been proposed as a method to shed light on the two production mechanisms~\cite{Mrowczynski:2019yrr}, coalescence versus thermal freeze-out. In that study, kaon-deuteron and proton-deuteron femtoscopy correlations were considered as probes of the deuteron formation process.

In fact, correlations involving a deuteron and a kaon or a proton have been successfully measured by several experimental collaborations in p-p and ion-ion collisions. In particular, proton-deuteron femtoscopy has been addressed in high-multiplicity p-p collisions at $\sqrt{s}=13$ TeV by the ALICE collaboration~\cite{ALICE:2023bny}, in Au-Au collisions at $\sqrt{s_{NN}}=3$ GeV by the STAR experiment~\cite{STAR:2024zvj}, and in Ag-Ag collisions at $\sqrt{s_{NN}}=2.55$ GeV by the HADES experiment~\cite{Stefaniak:2024eux}. In the case of p-p collisions, a three-body description has been shown to describe the $pd$ correlation function with remarkable accuracy~\cite{Viviani:2023kxw}. Alternatively, a two-body interaction framework that incorporates several partial waves can also reproduce the data rather well~\cite{Rzesa:2024oqp,Torres-Rincon:2024znb}, and it has also been successfully applied to Au-Au data from STAR~\cite{Torres-Rincon:2024znb}. Nonetheless, the two-body description might present conceptual issues, as raised in Refs.~\cite{Mrowczynski:2019yrr,Mrowczynski:2025qys}. 

Regarding kaon-deuteron femtoscopy, the $K^+d$ system has been analyzed in high-multiplicity p-p collisions at $\sqrt{s}=13$ TeV by the ALICE collaboration~\cite{ALICE:2023bny}, and in Pb-Pb collisions at $\sqrt{s_{NN}}=5.02$ TeV across different centralities in Ref.~\cite{Rzesa:2024dru} and in W. Rzesa's Ph.D. thesis~\cite{Rzesa:2024nra_thesis}. For the opposite charge channel, $K^-d$, no results have yet been published in p-p collisions by ALICE; however, femtoscopy analyses have been performed in PbPb collisions at $\sqrt{s_{NN}}=5.02$ TeV for several centralities~\cite{Rzesa:2024dru,Rzesa:2024nra_thesis}.

Using the experimental $K^+d$ correlation data in~\cite{ALICE:2023bny}, Ref.~\cite{VazquezDoce:2024nye} has recently investigated the possible time delay in deuteron production. Their findings suggest a preference for early deuteron formation via coalescence, rather than late direct production.

Since a two-body description of the kaon-deuteron systems appears to be physically well motivated~\cite{Mrowczynski:2019yrr,Mrowczynski:2025qys}, any femtoscopy analysis of $K^\pm d$ correlations should rely on a well-constrained interaction potential between deuterons and kaons. To date, theoretical studies and predictions of $K^\pm d$ femtoscopy have been predominantly based on the Lednick\'y-Lyuboshitz (LL) formalism~\cite{Lednicky:1981su,Lednicky:2005tb}, which allows modeling the correlation function using scattering lengths. Several published values of the $K^- d$ scattering length, plus some (unpublished) values of the $K^+ d$ scattering length, have been used in the most recent studies of kaon-deuteron femtoscopy~\cite{Rzesa:2024dru,Rzesa:2024nra_thesis}. However, a more complete picture of their interaction is still lacking in femtoscopy studies. 

In this work, we present a unified theoretical framework for kaon-deuteron femtoscopy based on chiral effective interactions, combined with the Faddeev equations. Using this approach, we compute the $T$ matrices for both $K^-d$ and $K^+ d$ systems using the same microscopic description. These scattering amplitudes, together with the effect of the Coulomb interaction, are used to reconstruct the pair wave functions and then input them into the Koonin-Pratt (KP) equation~\cite{Koonin:1977fh,Pratt:1990zq} to calculate the corresponding correlation functions. Our analyses will cover both small (p-p) and large (Pb-Pb) collision volumes, by selecting different source sizes in the formalism. Thus, we will provide a unified treatment of kaon-deuteron correlations across different experimental setups. 

The use of chiral effective theory in femtoscopy is supported by recent calculations of the $K^- p$ interaction~\cite{Kamiya:2019uiw,Encarnacion:2024jge} (also for $K^-\Lambda$ pairs in~\cite{Sarti:2023wlg}). Here we will combine the $K^-p$ case (along with its coupled channels), dominated by the $\Lambda(1405)$ resonance, together with the $K^- n$ interaction to construct the $K^- d$ scattering wave function. This will be achieved within the impulse approximation (IA) and the fixed-center approximation (FCA) of the Faddeev equations. Analogously, we will also work out the $K^+p$ and $K^+n$ interactions within the same formalism to address the $K^+d$ wave function. This approach goes beyond previous calculations that rely solely on the use of scattering lengths within the LL approximation. As will be shown, several effects---particularly those arising from the coupled-channel dynamics---are essential to an accurate description of the correlation functions via the KP formalism and cannot be captured within the simplified LL model.

The paper is organized as follows. In Sec.~\ref{sec:corr} we present our approach for obtaining the correlation functions of $K^-d$ and $K^+ d$ pairs from their respective wave functions. These are derived from the corresponding $K^-d$ and $K^+d$ amplitudes, which are obtained  within two approximations (IA and FCA), as detailed in Sec.~\ref{sec:ampl}. Our results for the $K^-d$, $K^+d$ amplitudes and scattering lengths, for the energy shift and width of kaonic deuterium and for the correlation functions of $K^-d$ and $K^+ d$ pairs
are discussed in Sec.~\ref{sec:results}, where the latter are also compared with preliminary available experimental data. A summary and some conclusions are given in Sec.~\ref{sec:conclusions}.

\section{Correlation function}
\label{sec:corr}
In the present work, the correlation function of the kaon-deuteron system will be taken 
within a two-body approach and it is given by the Koonin-Pratt expression \cite{Koonin:1977fh,Pratt:1990zq},
\begin{equation}
    C(k) = \int S_{12}( r) \left| \Psi( \bm{k};\bm{r}) \right|^2 d^3r \ ,
    \label{eq:corr}
\end{equation}
where $S_{12}( r)$ is the source function of the pair and $\Psi( \bm{k}; \bm{r})$ is the relative two-body wave function depending on the relative coordinate $\bm{r}$ and the relative momentum $
\bm{k}=(M_d\, \boldsymbol{p_K} - m_K \, \boldsymbol{p_d})/(m_K + M_d) $, where $M_d$ and $m_K$ are de deuteron and kaon masses, respectively.
As shown in Ref.~\cite{Mrowczynski:2025qys}, the three-body correlation approach reduces to the two-body expression of Eq.~(\ref{eq:corr}) if the total wave function factorizes into the deuteron one times the kaon-deuteron relative function $\Psi$, an assumption that appears reasonable given the distinguishability of the kaon with respect to the nucleons of the deuteron. 

The source function accounts for the probability per unit volume of finding a kaon-deuteron pair at a certain relative distance $r$ at freeze-out. It is usually taken as a Gaussian-type shape \cite{Fabbietti:2020bfg},
\begin{equation}
 S_{12}(r)=\frac{1}{(2\pi R_{Kd}^2)^{3/2}}{\rm e}^{-\frac{r^2}{2R_{Kd}^2}} \ ,
\end{equation}
where the relative size $R_{Kd}$ would be given by $R_{Kd}=\sqrt{R_K^2+R_d^2}$ if individual Gaussian distributions with sizes $R_K$ and $R_d$ for the kaon and deuteron, respectively, were assumed. Usually, the size is either taken from transverse mass scaling arguments \cite{ALICE:2020ibs} or assumed to be a free parameter fitted to the correlation function together with the low-energy scattering parameters of the pair interaction \cite{Rzesa:2024nra_thesis}.

Since the kaon ($K^-$ or $K^+$) and the deuteron ($d$) are charged particles, the wave function must include the effect of both the strong and electromagnetic interactions. As specified in the next section, the strong interaction is only considered in s-wave, an assumption that is reasonable for the low relative momenta of up to $~250$~MeV explored in the present work. With this assumption, the wave-function can be written as:
\begin{equation}
    \Psi( \bm{k};\bm{r}) =  \Phi^{\rm{C}}( \bm{k};\bm{r}) - \Phi_{0}^{\rm C}(kr) +  \Psi_0(k;r)    \ , \label{eq:wavefull}
\end{equation}
where $\Phi^{\rm{C}} ( \bm{k};\bm{r})$ is the complete Coulomb wave function and $\Phi_{0}^{\rm C} (kr)$ its $L=0$ component~\cite{joachain1975quantum}, whereas $\Psi_0$ is the kaon-deuteron wave function containing the effect of the strong and Coulomb forces in s-wave. With the above decomposition Eq.~(\ref{eq:corr}) becomes:
\begin{eqnarray}
C (k)&=&\int d^3r  S_{12}(r) \ |\Phi^{\rm C}(\boldsymbol{k};\boldsymbol{r}) |^2 \label{eq:corr2} \\
&+&\int 4\pi r^2\, dr  S_{12}(r) \left( | \Psi_0(k; r) |^2 - | \Phi^{\rm C}_{0} (k\, r ) |^2 \right) .
\nonumber
\end{eqnarray}

The wave function will obtained from the kaon-deuteron amplitude, $T_{Kd}$, by solving the Lippmann-Schwinger equation which in s-wave~\cite{joachain1975quantum},
\begin{eqnarray}
    \Psi_0(k; r) &=& j_0(k r) \label{eq:wf1} \\
    &+&
    \int_0^\infty \frac{d^3q}{(2\pi)^3} \frac{M_d}{E_d} \frac{1}{2\omega_{K} }\frac{T_{Kd}(q,k;\sqrt{s_{Kd}}) \, j_0(q r)}{\sqrt{s_{Kd}}-E_d -\omega_K + {\rm i}\eta } \ ,
    \nonumber
\end{eqnarray}
where $j_0(qr)$ is the spherical Bessel function, $\omega_K=\displaystyle\sqrt{q^2 + m_K^2}$ the kaon energy, $E_d$ the deuteron energy and $\displaystyle\sqrt{s_{Kd}}$ the energy of the kaon-deuteron pair in its center-of-mass (c.m.) frame.

The kaon-deuteron scattering amplitude $T_{Kd}$ could, in principle, be built from the solution of the corresponding Bethe-Salpeter (or Lippmann-Schwinger) equation, schematically written as
\begin{equation}
    T_{Kd} = V_{Kd}+ V_{Kd} \, G_{Kd} \, T_{Kd} 
    % = (V^{\rm C}_{Kd}+V^{\rm S}_{Kd})+ (V^{\rm C}_{Kd}+V^{\rm S}_{Kd}) \, G_{Kd} \, T_{Kd}
    \ ,
   \label{eq:BS_T} 
\end{equation}
employing the combined Coulomb and strong bare potentials, namely $V_{Kd}= V^{\rm C}_{Kd} + V^{\rm S}_{Kd}$.  While the Coulomb potential is well known, the strong $Kd$ potential could be phenomenologically built as an appropriate function with parameters fitted to data of $Kd$ scattering, following a similar strategy to that in our proton-deuteron work \cite{Torres-Rincon:2024znb}.  Since $Kd$ data are not available, a phenomenological potential $V_{Kd}$ cannot be adjusted. Instead, we follow a different strategy consisting of building the strong $Kd$ scattering amplitude $T^{\rm S}_{Kd}$ microscopically, i.e. deriving it from elementary $K^-N$ and $K^+N$ amplitudes ($N=p$ or $n$) that reproduce the scattering observables and are obtained within a chiral unitary formalism (see next section for details).

The use of a strong scattering amplitude $T_{Kd}$ for the $Kd$ problem, instead of a kernel potential $V_{Kd}$, prevents us from incorporating the additional Coulomb interaction employing Eq.~(\ref{eq:BS_T}). Alternatively, we will incorporate the Coulomb effects in a separable way as
\begin{equation}
 T_{Kd} \sim T_{Kd}^{\rm C} +  T_{Kd}^{\rm S} \ ,  
 \label{eq:T_sep}
\end{equation}
which amounts to ignoring the interference terms involving the strong and electromagnetic interactions that are present in the full $Kd$ amplitude. Nevertheless,  for the more sensitive case of $K^-d$ scattering, we have also employed the alternative approach based on the solution of the Schr\"odinger equation with appropriate potentials \cite{tobepublished}, finding the separable approximation to induce differences in the correlation function of at most 10\% with respect to the full treatment. Inserting the separable form into Eq.~(\ref{eq:wf1}), we obtain
\begin{widetext}
    \begin{align}
        \Psi_0(k; r) & = j_0(q r)  +
    \int_0^\infty \frac{d^3q}{(2\pi)^3} \frac{M_d}{E_d} \frac{1}{2\omega_{K} }\frac{T^{\rm C}_{Kd}(q,k;\sqrt{s_{Kd}}) \, j_0(q r)}{\sqrt{s_{Kd}}-E_d -\omega_K + {\rm i}\eta }  +   \int_0^\infty \frac{d^3q}{(2\pi)^3} \frac{M_d}{E_d} \frac{1}{2\omega_{K} }\frac{T^{S}_{Kd}(q,k;\sqrt{s_{Kd}}) \, j_0(q r)}{\sqrt{s_{Kd}}-E_d -\omega_K + {\rm i}\eta } \ \nonumber \\ 
   & =  \Phi^{\rm C}_{0} (k\, r )  + \Delta \varphi_0^{\rm S}(k;r) \ ,
 \end{align}
 \end{widetext}
where the Bessel function and the first integral on the right-hand side combine into the Coulomb wave function in s-wave,  $\Phi^{\rm C}_0$, shown after the last equality. The remaining term is the correction induced by the strong interaction, $\Delta \varphi_0^{\rm S}$ and, in practical terms, it is obtained from:
\begin{eqnarray}
\Delta \varphi_0^{\rm S}(k;r) &=&
    T_{Kd}^{\rm S}(\sqrt{s_{Kd}})  \\  &\times& \int_0^\Lambda \frac{d^3q}{(2\pi)^3} \frac{M_d}{E_d} \frac{1}{2\omega_{K} }\frac{j_0(q r)}{\sqrt{s_{Kd}}-E_d -\omega_K + {\rm i}\eta }\ ,   \nonumber
\end{eqnarray}
where, consistent with its derivation, the kaon-deuteron amplitude $T^{\rm S}_{Kd}$ is factorized on-shell out of the integral. The integral is extended up to a cutoff momentum $\Lambda$ for which we take a value of 1000 MeV.

\section{\mbox{\boldmath $K^-d$} and \mbox{\boldmath $K^+d$} amplitudes}
\label{sec:ampl}

The study of the $K^- d$ interaction has attracted a lot of attention for more than 50 years, since the pioneer Faddeev calculation of the low energy $K^-d$ cross sections~\cite{Hetherington:1965zza}, later improved by several authors by including explicitly the coupling to hyperonic $\pi N Y$ ($Y=\Lambda,\Sigma$) channels~\cite{Toker:1981zh,Torres:1986mr}, which were found to have little effect.

\begin{figure*}[ht]
    \centering   \includegraphics[width=0.85\linewidth]{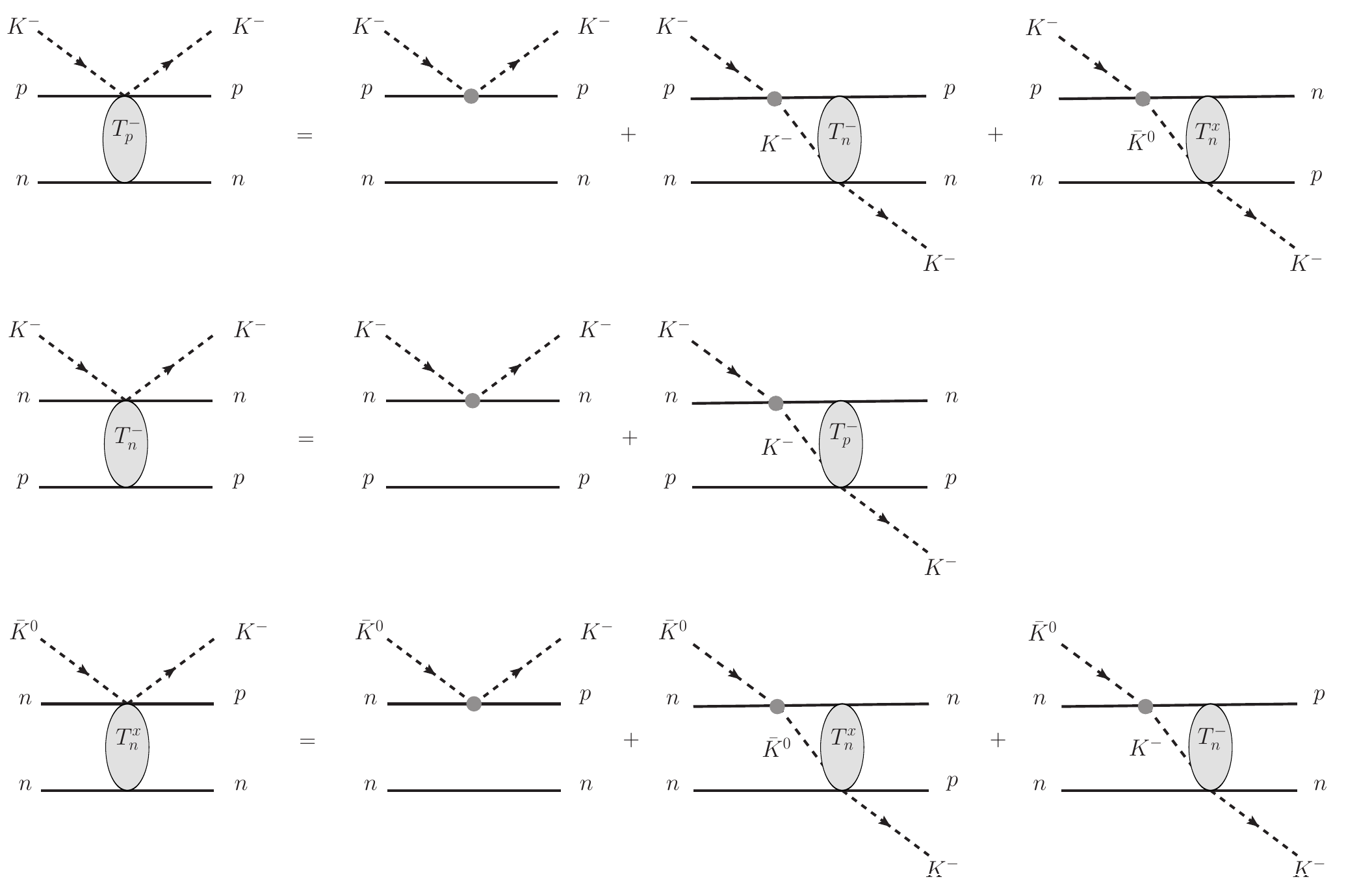}
    \caption{Diagrams contributing to the $K^-d$ Faddeev equations in the fixed-center approximation. The three rows correspond to the Faddeev partitions denoted as $T_p^-$ (top), $T_n^-$ (middle) and $T_n^\mathrm{x}$ (bottom).}
    \label{fig:faddeev_kmD}
\end{figure*}

Soon after the development of the unitarized chiral effective approach \cite{Kaiser:1995eg,Oset:1997it}, which permits obtaining the $\bar K N$ elastic and inelastic scattering amplitudes incorporating dynamically the effects of the subthreshold resonance $\Lambda(1405)$, the $K^- d$ scattering length was derived within the so-called FCA to the Faddeev equations~\cite{Kamalov:2000iy}, consisting in disregarding the recoil effects of the nucleons within the deuteron when they interact with the $K^-$. 
This simplification allows us to obtain a straightforward determination of the $K^- d$ scattering length, $A_{K^- d}$, 
in terms of $\bar{K}N$ two-body amplitudes at threshold and, via the corrected Deser formula~\cite{Meissner:2005bz}, have an estimation of the energy shift and width of the $1s$ kaonic deuterium state.  This fact was exploited in~\cite{Meissner:2006gx} to estimate the feasibility of extracting the elementary $\bar KN$ scattering lengths from combining the available measurements of the shift and width of the $1s$ kaonic hydrogen state~\cite{DEAR:2005fdl} with synthetic data for the kaonic deuterium one, soon to become available \cite{SIDDHARTA-2_Sgaramella,SIDDHARTA-2:2025ulg}. The analysis was redone~\cite{Doring:2011xc} in view of the more precise kaonic hydrogen data provided by SIDDHARTA~\cite{SIDDHARTA:2011dsy} and compatible with that from KEK~\cite{Iwasaki:1997wf}, leading to a prediction of $A_{K^- d}$ with around 20\% and 30\% uncertainties for the real and imaginary parts, respectively, tied to the uncertainties of the elementary $K^-p$ amplitudes extracted from scattering and kaonic hydrogen data. 
We note that the shift and width of the kaonic atom state can be directly obtained either from a two-body reduced~\cite{Barrett:1999cw} or a three-body Schr\"odinger equation~\cite{Hoshino:2017mty}, which show that the FCA estimates of $A_{K^- d}$ may induce differences in the estimation of the energy shift and/or width of up to 30\%.

Sophisticated Faddeev calculations of the $K^-d$ problem employing elementary low-energy expansion amplitudes~\cite{Deloff:1999gc}, phenomenological potentials~\cite{Shevchenko:2011ce} and chirally inspired interactions \cite{Bahaoui:2003xb} have been revisited~\cite{Mizutani:2012gy,Shevchenko:2012np,Shevchenko:2014uva} in view of the new kaonic hydrogen measurement~\cite{SIDDHARTA:2011dsy}. The comparison between the full Faddeev and FCA results in these works, as well as the instructive comparison discussed in~\cite{Gal:2006cw}, lead to the conclusion that the FCA approximation can differ from the full Faddeev calculation in 10-30\%, well within the uncertainties tied to the different elementary model interactions employed in the literature.

\begin{figure*}[ht]
    \centering   \includegraphics[width=0.85\linewidth]{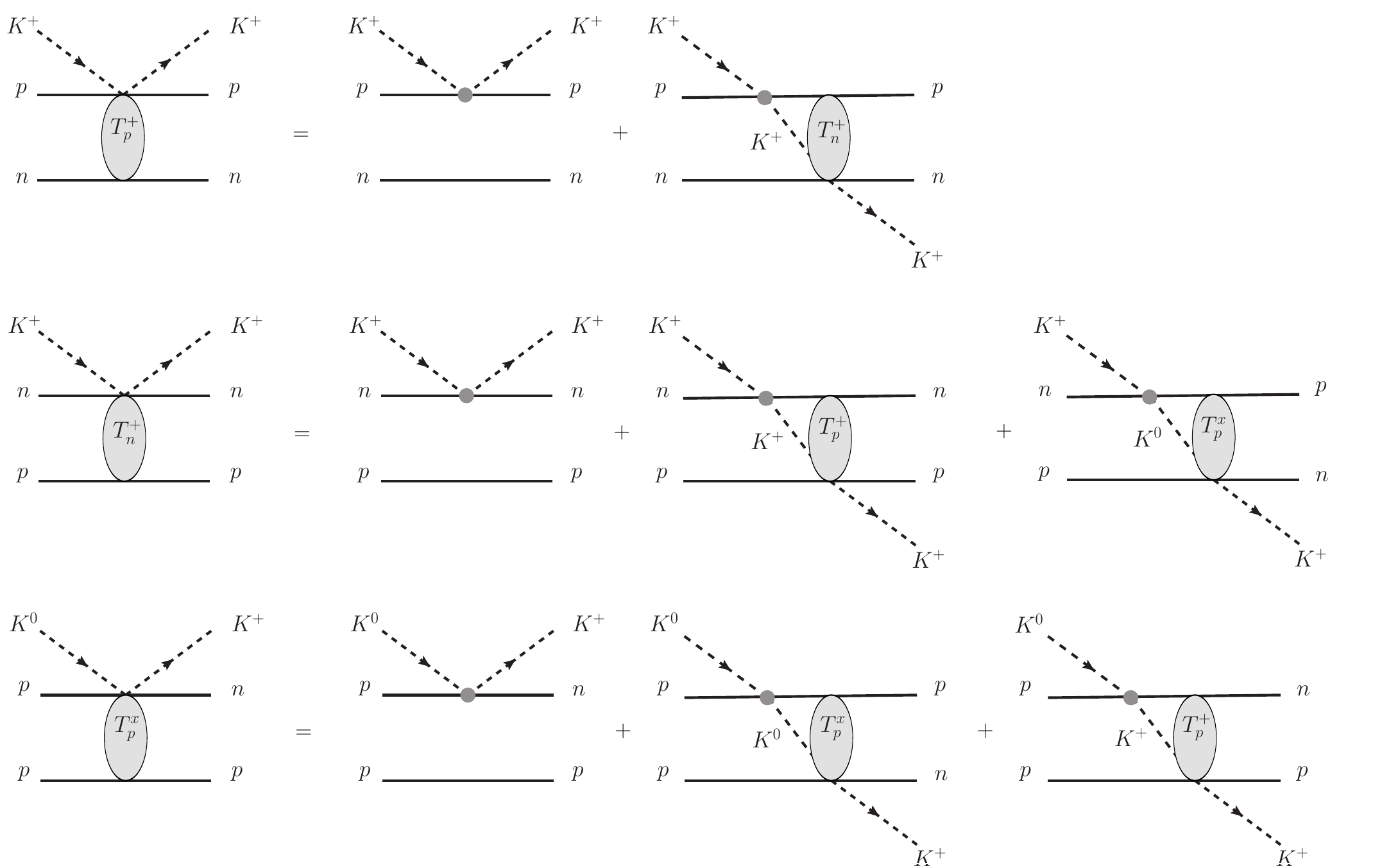}
        \caption{Diagrams contributing to the $K^+d$ Faddeev equations in the fixed-center approximation. The three rows correspond to the Faddeev partitions denoted as $T_p^+$ (top), $T_n^+$ (middle) and $T_p^\mathrm{x}$ (bottom) scatterings.}
    \label{fig:faddeev_kpD}
\end{figure*}

In view of the above discussion, the $K^-d$ amplitude, $T_{K^- d}$, is derived in the present exploratory work from the FCA approach to the Faddeev equations, which are displayed diagrammatically in Fig.~\ref{fig:faddeev_kmD}. For completeness, we will compare our results with those of the simpler IA, represented by the sum of the first diagram on the right-hand side of the first and second rows in the figure. This approximation considers single-kaon scattering off a stationary nucleon, either the proton or the neutron, assuming the other to be a spectator. Similarly, the $K^+ d$ amplitude, $T_{K^+ d}$, will be obtained from the diagrams of Fig.~\ref{fig:faddeev_kpD}. The gray circles in Figs.~\ref{fig:faddeev_kmD} and \ref{fig:faddeev_kpD} denote two-body s-wave amplitudes, which we obtain within the chiral unitary model of Ref.~\cite{Oset:1997it} adopting the updated version of Ref.~\cite{Jido:2002zk}. The resulting scattering lengths at the corresponding physical thresholds are displayed in Table~\ref{tab:scatKN}. 

\begin{table}[ht]
    \centering
    \begin{tabular}{c|c}
    \hline
    \hline
              & \\[-3mm]
     $a_{K^- p}$ [fm] &    $-0.94 + {\rm i} 1.00$  \\
%           $a_{\bar{K}^0 n}$ [fm] &  $ -0.40 + {\rm i} 1.18$ \\
     ($\sqrt{s}=1431.95$ MeV) &  \\
      $a_{K^- n}$ [fm] &   $0.45 + {\rm i} 0.78$ \\
       ($\sqrt{s}=1433.24$ MeV)  & \\
      $a_{\bar{K}^0 n}$ [fm] &  $ -0.58 + {\rm i} 1.12$ \\
       ($\sqrt{s}=1437.24$ MeV)  & \\
     \hline
          & \\[-3mm]
 $a_{K^+ p}$ [fm] &       $-0.26$   \\
  ($\sqrt{s}=1431.95$ MeV) &    \\
  $a_{K^+ n}$ [fm] &  $-0.13$   \\
    ($\sqrt{s}=1433.24$ MeV) &   \\   
    $a_{K^0 p}$ [fm] &  $-0.13$   \\
  ($\sqrt{s}=1435.94$ MeV) &   \\   
       \hline
      \hline
    \end{tabular}
    \caption{$K^-N$ and $K^+N$ scattering lengths of the elementary interaction model employed \cite{Oset:1997it,Jido:2002zk}}
    \label{tab:scatKN}
\end{table}

The $K^-d$ amplitude in the IA,  $T_{K^-d}^{\mathrm{IA}}$, is simply given in terms of the $t$-matrices that describe the $K^-N$ scattering on the proton $t_{K^-p, K^-p}$ and on the neutron $t_{K^-n, K^-n}$ \cite{Kamalov:2000iy}
    \begin{equation}
        T_{K^-d}^{\mathrm{IA}}(k',k)=[t_{K^-p, K^-p}(k',k) + t_{K^-n, K^-n}(k',k)]F_d(q) \ ,
        \label{eq:T_IA}
    \end{equation}
    where $\bm{q}=(\bm{k} - \bm{k}\,^\prime)/2$ is half the momentum transferred to the deuteron, built from initial and final kaon momentum $\bm{k}$ and $\bm{k}\,^\prime$, respectively, and
    \begin{equation}
        F_d(q)=\int  \mathrm{d}^3r \ e^{-{\rm i} \bm{q}\cdot \bm{r}}|\phi_d(\bm{r}\,)|^2
    \end{equation}
    is the elastic deuteron form factor normalized to unity at $q=0$, with $\phi_d(\bm{r})$ being the deuteron wave function. This wave function is divided into the usual s- and d-wave components $u(r)$ and $w(r)$, respectively, for which we take the analytical expression given in Ref.~\cite{Mahlein:2023fmx} with parameters fitted to the realistic deuteron wave function obtained with the Argonne V18 interaction \cite{Wiringa:1994wb}.  
    We have redone all of our calculations with the Paris deuteron wave function~\cite{Lacombe:1981eg} employed in Ref.~\cite{Kamalov:2000iy}. We find the correlation functions obtained with either choice of the deuteron wave function to be barely distinguishable.

For the $K^+d$ scattering amplitude in IA we similarly obtain the same expression as in Eq.~(\ref{eq:T_IA}), but replacing the elementary $t$-matrices by $t_{K^+p, K^+p}$ and $t_{K^+n, K^+n}$.

Turning to the calculation in the FCA, the $K^- d$ amplitude is given by
    \begin{equation}
        T_{K^-d}^\mathrm{FCA} = T^-_p + T^-_n,
    \end{equation}
    where the amplitudes $T^-_p$ and $T^-_n$ correspond to the Faddeev partitions that contain the processes in which the $K^-$ collides first on a proton and a neutron, respectively. This is seen in the diagrams of Fig.~\ref{fig:faddeev_kmD}, which represent the  following system of coupled equations:
\begin{eqnarray}
    T^-_p &=& t_{K^-p,K^-p} + t_{K^-p,K^-p} G_0 T^-_n - t_{K^-p,\bar{K}^0 n} G_0 T^\mathrm{x}_n \nonumber \\
    T^-_n &=& t_{K^-n,K^-n} + t_{K^-n,K^-n} G_0 T^-_p 
        \label{eq:TKm}
        \\
    T^\mathrm{x}_n &=& t_{\bar{K}^0n,K^-p} - t_{\bar{K}^0n,\bar{K}^0n} G_0 T^\mathrm{x}_n + 
    t_{\bar{K}^0n,K^-p} G_0 T^-_n  \ ,  \nonumber
\end{eqnarray}
    where $T_n^\mathrm{x}$ is the Faddeev partition that describes the $\bar{K}^0nn \to K^-pn$ transition, including the multiple scattering in the intermediate states, and $G_0$ is the free kaon propagator given by    
    \begin{equation}
        G_0 (q^0) = \int\frac{\mathrm{d}^3 q}{(2\pi)^3}\frac{F_d(q)}{(q^0)^2-\bm{q}^2-m_K^2+i\eta} \ ,
        \label{eq:prop0}
    \end{equation} 
    where $m_K$ is the mass of the corresponding propagating kaon ($K^-$ or ${\bar K}^0$).
    Note that the $K^- p \to \bar{K}^0 n$ exchange amplitude appearing in the diagrams of Fig.~\ref{fig:faddeev_kmD} transforms the initial proton of the deuteron into a neutron and, therefore, the intermediate state is $\bar{K}^0 nn$. Since the FCA considers the two nucleons as static, the two-neutron wave-function of this intermediate state is not modified and is taken to be the same as the deuteron one. This is a limitation of the FCA that will need to be addressed in the future with a more sophisticated approach.

In the set of Eqs.~(\ref{eq:TKm}) note the minus sign in the terms that involve the exchange of the $pn$ pair into a $np$ one. The solution of these equations is
\begin{widetext}
    \begin{equation}
  T_{K^-d}^\mathrm{FCA} = \frac{T_{K^-d}^\mathrm{IA}  + \left(2t_{K^-p, K^-p} t_{K^-n, K^-n} -\tilde{t}_{K^-p, \bar{K}^0n}^2\right) G_0  
    - 2\tilde{t}_{K^-p, \bar{K}^0n}^2 t_{K^-n, K^-n} G_0^2}
    {1 - t_{K^-p, K^-p} t_{K^-n, K^-n} G_0^2 + \tilde{t}_{K^-p, \bar{K}^0n}^2 t_{K^-n, K^-n} G_0^3}
    \label{eq:Kmd}
    \end{equation}
\end{widetext}%    \begin{eqnarray}
%        T_{K^-d}^\mathrm{FCA} =&& \frac{1}{\text{den}}\left[T_{K^-d}^\mathrm{IA}  \right. \nonumber \\
%        &+& \left(2t_{K^-p \to K^-p} t_{K^-n \to K^-n} -\tilde{t}_{K^-p\to \bar{K}^0n}^2\right) G_0  \nonumber \\
%        &-& 2\tilde{t}_{K^-p\to \bar{K}^0n}^2 t_{K^-n \to K^-n} G_0^2 \Big]  \ ,
%\end{eqnarray}
with  
%\begin{eqnarray}
%    \text{den}=&&1 - t_{K^-p \to K^-p} t_{K^-n \to K^-n} G_0^2 \nonumber \\
%    &+& \tilde{t}_{K^-p\to \bar{K}^0n}^2 t_{K^-n \to K^-n} G_0^3 \ ,
%    \end{eqnarray}  
%    and          
    \begin{equation}
        \tilde{t}_{K^-p, \bar{K}^0n} = \frac{t_{K^-p, \bar{K}^0n}}{\sqrt{1+t_{\bar{K}^0n, \bar{K}^0n}G_0}}.
    \end{equation}
The expression in Eq.~(\ref{eq:Kmd}) is an extension of the scattering length formula, given in Ref.~\cite{Kamalov:2000iy}, to higher energies and therefore suitable for the evaluation of the correlation function at finite momentum values. Note that this extension was also employed in Ref.~\cite{Oset:2012gi}, but we have implicitly implemented the form factor into the single-scattering terms collected in $T^\mathrm{IA}_{K^-d}$, so that the IA limit is recovered when taking $G_0 \to 0$, i.e. ignoring the rescattering effects.

Similarly, we can express the $T$-matrix of the $K^+d$ interaction as
    \begin{equation}
        T_{K^+d}^\mathrm{FCA} = T^+_p + T^+_n,
    \end{equation}
    where $T^+_p$ and $T^+_n$ correspond to the Faddeev partitions describing the interaction between $K^+$ and the deuteron starting with a first collision on a proton and a neutron, respectively. This is seen in the diagrams of Fig.~\ref{fig:faddeev_kpD}, which represent the following system of coupled equations:
\begin{eqnarray}
    T^+_p &=& t_{K^+p,K^+p} + t_{K^+p,K^+p} G_0 T^+_n   
        \label{eq:TKp} \\
    T^+_n &=& t_{K^+n,K^+n} + t_{K^+n,K^+n} G_0 T^+_p - t_{K^+n,{K}^0 p} G_0 T^\mathrm{x}_p  \nonumber \\
    T^\mathrm{x}_p &=& t_{{K}^0p,{K}^+n} - t_{{K}^0p,{K}^0p} G_0 T^\mathrm{x}_p + 
    t_{{K}^0p,K^+n} G_0 T^+_p  \ , \nonumber  \\
    \nonumber
\end{eqnarray}
where the $T_p^\mathrm{x}$ Faddeev partition describes the $K^0pp \to K^+np$ transition that includes the multiple scattering in the intermediate states. The solution is
\begin{widetext}
    \begin{equation}
       T_{K^+d}^\mathrm{FCA} =  \frac{  T_{K^+d}^\mathrm{IA}   +  \left(2t_{K^+p, K^+p} t_{K^+n, K^+n} - \tilde{t}_{K^+n, K^0p}^2\right) G_0 
         - 2 \tilde{t}_{K^+n, K^0p}^2 t_{K^+p, K^+p} G_0^2}
        {1 - t_{K^+p, K^+p} t_{K^+n, K^+n} G_0^2 +\tilde{t}_{K^+n, K^0p}^2 t_{K^+p, K^+p} G_0^3}          \ ,      
    \end{equation}
\end{widetext}
with  
\begin{equation}
        \tilde{t}_{K^+n, K^0p} = \frac{t_{K^+n, K^0p}}{\sqrt{1+t_{K^0p, K^0p}G_0}} \ .
    \end{equation}

\section{Results} \label{sec:results}

\subsection{Scattering amplitudes and scattering length}

\begin{figure*}[h!]
        \centering
        \includegraphics[width=0.49\linewidth]{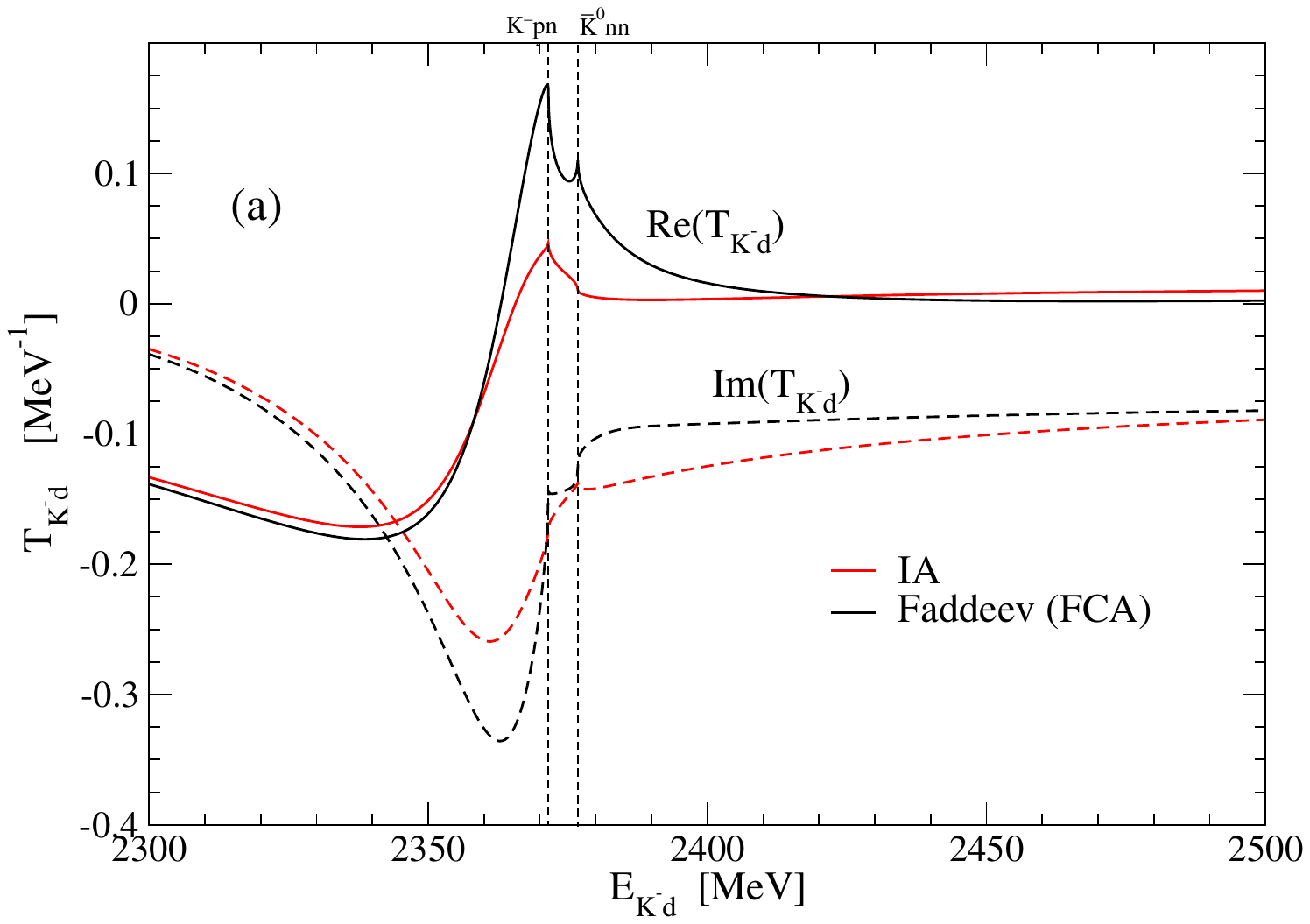}
    \hfill
        \includegraphics[width=0.49\linewidth]{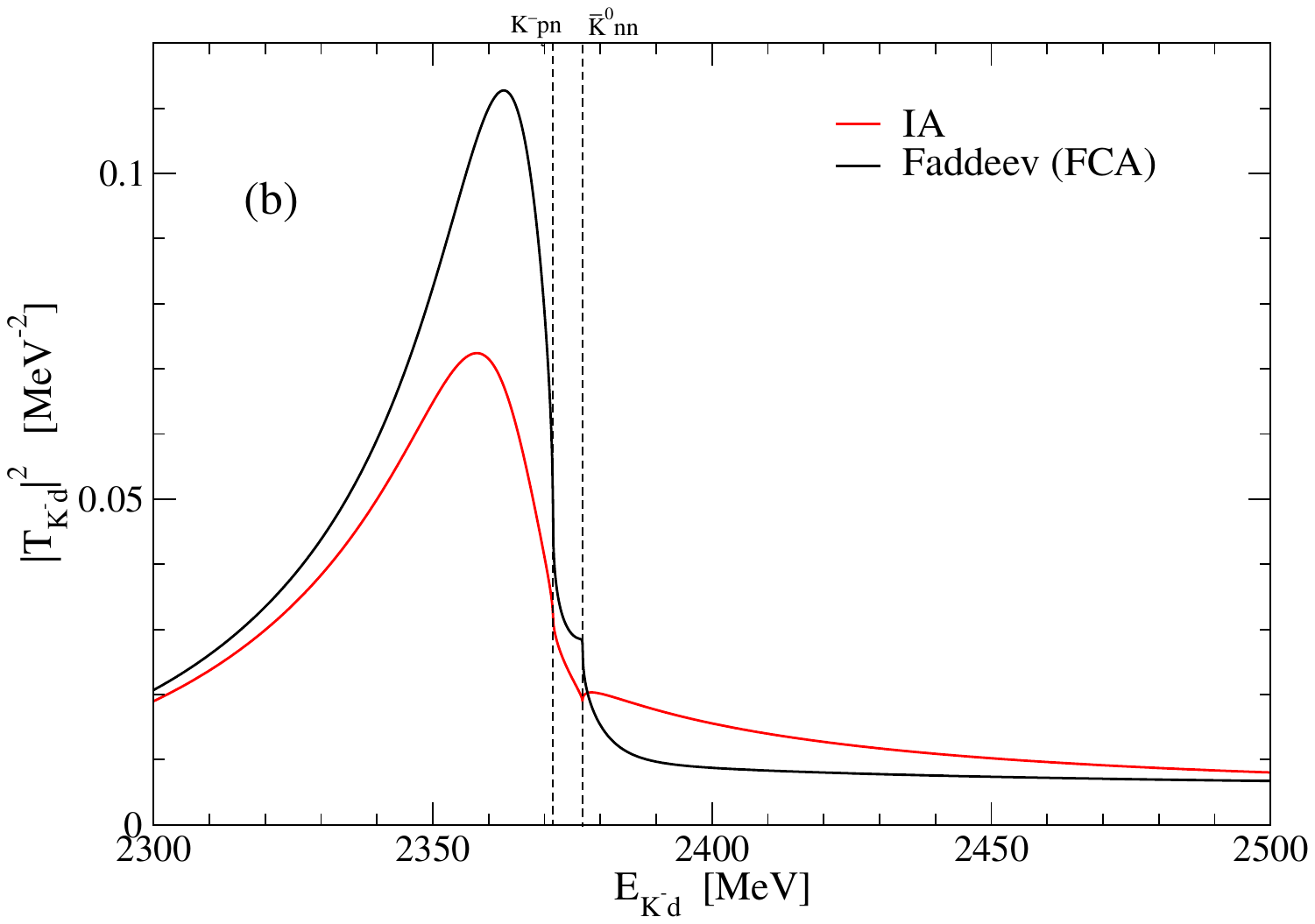}
    \caption{(a) Scattering amplitude $T_{K^- d}$ for $K^- d \to K^- d$ around and below the $K^- d$ threshold. The solid lines represent the real part and the dashed lines the imaginary part. The red lines correspond to the IA and the blue lines to the FCA. (b) Squared modulus of the scattering amplitude $|T_{K^- d}|^2$. The $K^-pn$ and $\bar{K}^0nn$ thresholds are indicated with vertical dashed lines.}
    \label{fig:t_kmd}
    \end{figure*}

\begin{figure*}[h!]
    \centering
        \includegraphics[width=0.49\linewidth]{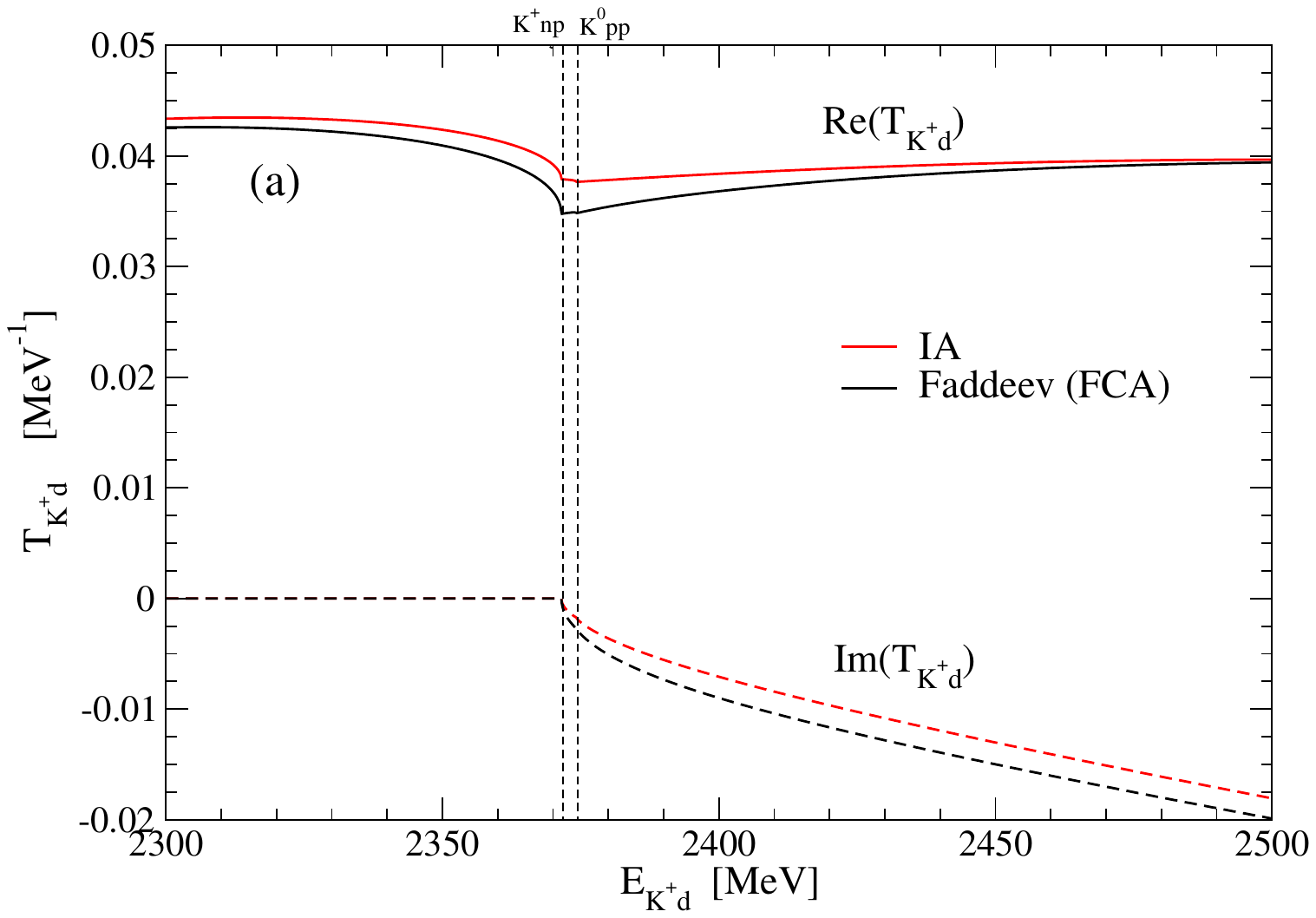}
    \hfill
        \includegraphics[width=0.49\linewidth]{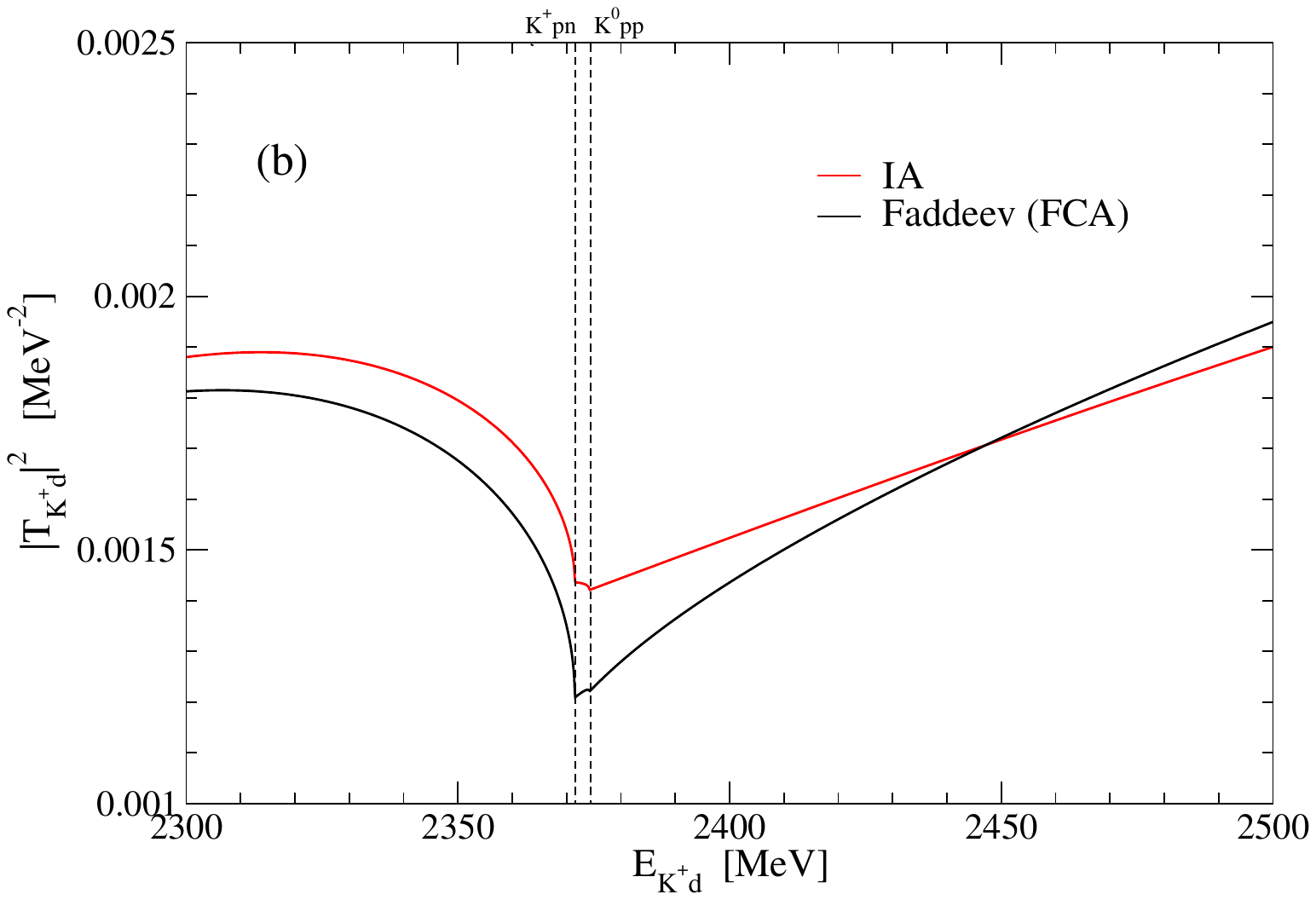}
    \caption{(a) Scattering amplitude $T_{K^+ d}$ for the process $K^+ d \to K^+ d$ around and below the $K^+ d$ threshold. The solid lines represent the real part and the dashed lines the imaginary part. The red lines correspond to the IA and the blue lines to the FCA. (b) Squared modulus of the scattering amplitude $|T_{K^+ d}|^2$. The $K^+ np$ and the $K^0pp$ thresholds are indicated with vertical dashed lines.}
    \label{fig:t_kpd}
    \end{figure*}

In Fig.~\ref{fig:t_kmd}  we represent the real and imaginary parts of the $T_{K^-d}$ amplitude (left panel) and its modulus squared (right panel) for the IA and FCA models studied in this work. Similarly, Fig.~\ref{fig:t_kpd} displays the same results for the $K^+d$ system. 

In the case of $K^-d$ scattering, the pronounced structure observed below the $K^-pn$ threshold (indicated by the vertical dashed-line on the left) is a reflection of the $\Lambda(1405)$ resonance present in the isospin $I=0$ ${\bar K}N$ elementary amplitude. The structure is noticeably enhanced when calculated using the FCA compared to the IA approach. This result is a clear evidence of the importance that the multi-scattering processes have in the dynamics of the  $\bar{K}NN$ system, essentially due to the strong energy-dependence of the ${\bar K}N$ elementary interaction dominated by the subthreshold $\Lambda(1405)$ and indicated by a very sizable negative real part of the $K^- p$ scattering length shown in Table~\ref{tab:scatKN}.

An important feature of the analysis is the appearance of cusp structures, that are better seen in the scattering amplitude shown on the left panel of Fig.~\ref{fig:t_kmd}. These nonanalytic features emerge at the opening of the $K^-pn$ and $\bar{K}^0nn$ channels, and their presence highlights the strong coupled channel effects and the rich energy dependence inherent in the three-body dynamics of the $K^-d$ system.

In contrast, the $K^+d$ scattering amplitude,  $T_{K^+d}$, shown in Fig.~\ref{fig:t_kpd} shows minimal differences between the FCA and IA results. The amplitude is practically structureless and has a moderate strength over the full energy range considered, which is a consequence of the relatively weak and repulsive elementary $KN$ interaction, absent of inelastic channels, as indicated by the $K^+N$ scattering lengths displayed in Table~\ref{tab:scatKN}. Correspondingly, the cusp structures at the $K^+np$ and $K^0pp$ thresholds in the $K^+d$ amplitude stand out less than those in the $K^-d$ case. 

We note that a series of recent papers~\cite{Encarnacion:2025lyf,Ikeno:2025bsx,Agatao:2025ckp} pointed out that, in order to have a {\it particle-cluster} amplitude that respects elastic unitarity and is therefore equivalent to that obtained from a Schr\"odinger equation for the interacting pair, one needs to improve the FCA amplitudes by considering the coherent propagation of the particle and the cluster (the $K^{\pm}$ and the deuteron, respectively, in our case). Employing isospin averaged amplitudes and ignoring the exchange terms, a procedure to implement the coherent propagation in the FCA is proposed in Ref.~\cite{Ikeno:2025bsx} and mathematically justified in Ref.~\cite{Agatao:2025ckp}. As isospin symmetry is slightly broken in our approach by the use of physical particle masses, and exchange processes are also incorporated, we cannot apply the procedure outlined in Refs.~\cite{Ikeno:2025bsx,Agatao:2025ckp} straightforwardly. However, we note that part of the coherent propagation might already be contained in the terms proportional to $G_0$ in Eqs.~(\ref{eq:TKm}) or (\ref{eq:TKp}), which in turn ensure the opening of the  unitary cut at the $K^{\pm} d$ two-body threshold. Moreover, since the $K^{\pm}pn$ three-body unitary cut starts only 2.22 MeV above the $K^{\pm} d$ one, we shift the amplitudes displayed in Figs.~\ref{fig:t_kmd} and \ref{fig:t_kpd} by this small amount to the left, which essentially opens the unitary cut at the  $K^{\pm} d$ threshold even in the impulse approximation. 
%With this prescription, the linear term in $k^*$ (momentum of the $K^-$ momentum in the $K^- d$ c.m. frame) of the inverse of the amplitude behaves as $-{\rm i} \alpha k^*$, with $\alpha \sim 0.7$ instead of 1, which falls within the uncertainties expected for the FCA. 
%}

The scattering length is related with the strong $K d$ scattering amplitude at threshold by
\begin{equation}
   A_{Kd}=-\frac{1}{4\pi}\frac{M_d}{m_{K}+M_d}T_{Kd} \ ,
\end{equation}
where $K$ may stand for the $K^-$ or the $K^+$ depending on the case one is considering. 
The scattering lengths obtained using the IA and FCA are presented in Table \ref{tab:scatlen_kd} and reveal strikingly different behaviors in the $K^-d$ and $K^+d$ systems.

In the case of $K^-d$, the large negative values of the real part are tied to a strong attractive interaction which gives rise to a subthreshold ${\bar K}NN$ resonance, in both the IA and the FCA. Note that the FCA predicts a substantially more negative real part of the scattering length compared to the IA result. The imaginary part is large and also affected by the type of approximation considered, with the FCA result being somewhat reduced compared to the IA one, suggesting that inelastic channels are somewhat suppressed when multiple scattering effects are properly accounted for. 

%{\color{blue} 

%These observations underline the importance of including re-scattering and three-body dynamics to achieve a more realistic description of the system.
Turning to the $K^+d$ system, it is found that both IA and FCA yield nearly identical scattering lengths. This is a reflection of the simpler underlying dynamics tied to a weaker interaction, predominantly elastic, and largely unaffected by higher-order corrections. 

    \begin{table}[h!]
        \centering
        \begin{tabular}{c c c}
        \hline
        \hline
            & $A_{K^-d}\, [\mathrm{fm}]$ & $A_{K^+d}\, [\mathrm{fm}]$ \\
            & $(\sqrt{s}=2369.290\,\mathrm{MeV})$ & $(\sqrt{s}=2369.290\,\mathrm{MeV})$ \\
        \hline
            IA & $-0.59+{\rm i} 2.15$ & $-0.47$ \\
            FCA & $-2.06+{\rm i} 1.77$ & $-0.43$ \\
        \hline
        \hline
        \end{tabular}
                \caption{Scattering lengths for $K^-d$ and $K^+d$.}
        \label{tab:scatlen_kd}
    \end{table}

In Table~\ref{tab:scatlen_kmd_comp} we provide a summary of selected previous results in the literature for $A_{K^-d}$. As can be seen, the data show a wide variation in both real and imaginary parts, with a consistent sign confirming the attractive nature of the interaction, with the presence of a ${\bar K}NN$ bound state originated from the $\Lambda(1405)$. Our final result in the FCA has the largest real part of the scattering length (in absolute value) and a similar imaginary part, compatible with other determinations. 

The dispersion of results reinforces the importance of experimental efforts that seek to provide information of the $K^- d$ interaction, such as the kaonic deuterium measurements of SIDDHARTA2 \cite{SIDDHARTA-2_Sgaramella,SIDDHARTA-2:2025ulg} and E57@JPARC \cite{J-PARCE57:2019adr}, or the femtoscopic analyses of ALICE \cite{ALICE:2023bny,Rzesa:2024dru,Rzesa:2024nra_thesis}.

\begin{table}[h!]
        \centering
        \begin{tabular}{c c c}
        \hline
        \hline
            Reference & Year & $A_{K^-d}\, [\mathrm{fm}]$ \\
        \hline
            Kamalov et al., \cite{Kamalov:2000iy} &2001 & -1.62 + {\rm i}1.91  \\
            Bahaoui et al., \cite{Bahaoui:2003xb} & 2003 & -1.80 + {\rm i}1.55  \\
            Shevchenko, \cite{Shevchenko:2011ce} (one pole) & 2011 & -1.49 + {\rm i}0.98  \\
            Shevchenko, \cite{Shevchenko:2011ce} (two pole) & 2011 & -1.57 + {\rm i}1.11  \\
            Oset et al., \cite{Oset:2012gi} & 2012 & -1.54 + {\rm i}1.82  \\
            Mizutani et al.,\cite{Mizutani:2012gy} & 2012 & -1.58 + {\rm i}1.37  \\
%            \cite{Revai:2012fx} (2012, one pole)            & & \\
%            \cite{Revai:2016muw}
            Hoshino et al., \cite{Hoshino:2017mty} & 2017 & -1.42 + {\rm i}1.60  \\            
            Liu et al., \cite{Liu:2020foc} & 2020 & -0.59 + {\rm i}2.70  \\
            Barnea et al., \cite{Barnea:2020dnj} & 2020 & -1.26 + {\rm i}1.41  \\
           \hline
        \hline
        \end{tabular}
                \caption{Scattering lengths of the $K^-d$ interaction obtained in previous calculations.}
        \label{tab:scatlen_kmd_comp}
    \end{table}

In Table~\ref{tab:scatlen_kpd_comp} we show two previous predictions for the scattering length of the $K^+d$ interaction. Both are unpublished results, presented in Ref.~\cite{Rzesa:2024nra_thesis} in the context of $K^+d$ femtoscopy. Our value is consistent with these results, showing the weak repulsive nature of the interaction.

\begin{table}[h!]
        \centering
        \begin{tabular}{c c c}
        \hline
        \hline
            Ref. & Year  & $A_{K^+d}\, [\mathrm{fm}]$  \\
        \hline
            Haidenbauer, \cite{Rzesa:2024nra_thesis} & 2025 & -0.54  \\
            Hyodo, \cite{Rzesa:2024nra_thesis} & 2025 & -0.47  \\
           \hline
        \hline
        \end{tabular}
                \caption{Scattering length of the $K^+d$ interaction obtained in previous calculations. Both results are quoted as private communications to the author of Ref.~\cite{Rzesa:2024nra_thesis}.}
        \label{tab:scatlen_kpd_comp}
    \end{table}

%\begin{table}[ht]
%    \centering
%    \begin{tabular}{c|c|c}
%%    \hline
%    \hline
%     & OR model & BCN model\\
%     \hline
%     & & \\[-3mm]
%     $a_{K^- p}$ [fm] & $-0.94 + {\rm i} 1.00$ & $-0.67 + {\rm i} 0.86$\\
%     ($\sqrt{s}=1431.95$ MeV) & & \\
%     $a_{K^- n}$ [fm] & $0.45 + {\rm i} 0.78$ & $0.55 + {\rm i} 0.60$ \\
%     ($\sqrt{s}=1433.24$ MeV) & & \\[1mm]
%     \hline
%          & & \\[-3mm]
%      $A_{K^- d}$ [fm]   &  &  \\
%      IA & $-0.59 + {\rm i} 2.15$ & $-0.15 + {\rm i} 1.76$\\
%      FCA  & $-2.06 + {\rm i} 1.77$ &  $-1.82 + {\rm i} 1.75$\\
%      \hline
%      \hline
%    \end{tabular}
%    \caption{$K^- n$, $K^- p$ and $K^- d$ scattering lengths}
%    \label{tab:scatKm}
%\end{table}

%\begin{table}[ht]
%    \centering
%    \begin{tabular}{c|c}
%%    \hline
%    \hline
%     & OR model \\
%     \hline
%          &  \\[-3mm]
%     $a_{K^+ p}$ [fm] & $-0.26$ \\
%     ($\sqrt{s}=1431.95$ MeV) &  \\
%     $a_{K^+ n}$ [fm] & $-0.13$  \\
%    ($\sqrt{s}=1433.24$ MeV) &  \\[1mm]
%    \hline
%         & \\[-3mm]
%      $A_{K^+ d}$ [fm]   &   \\
%      IA & $-0.47$ \\
%      FCA  & $-0.43$  \\
%%      \hline
%      \hline
%    \end{tabular}
%    \caption{$K^+ n$, $K^+ p$ and $K^+ d$ scattering lengths}
%    \label{tab:scatKp}
%\end{table}

\subsection{Kaonic deuterium}

The prospect of soon having the analysis of the SIDDHARTA2 collaboration providing a measurement of the energy shift, $\varepsilon_{1s}$, and width, $\Gamma_{1s}$, of the $1s$ level in kaonic deuterium \cite{SIDDHARTA-2_Sgaramella,SIDDHARTA-2:2025ulg} makes a prediction of these quantities timely. Within the present approach and employing the Deser-type formula including isospin-breaking effects \cite{Meissner:2006gx}
\begin{equation}
\varepsilon_{1s} - {\rm i}\frac{\Gamma_{1s}}{2} =
- 2 \alpha^3 \mu_r^2 A_{Kd} \left\{1 -2\alpha \mu_r A_{Kd} (\ln\alpha -1) + \dots\right\} \ ,
\end{equation}
where $\alpha$ is the fine structure constant and $\mu_r$ the reduced kaon-deuteron mass, we obtain the results shown in Table~\ref{tab:Kdatom} within the two approximations explored in this work.
\begin{table}[ht]
    \centering
    \begin{tabular}{c|c c}
    \hline
    \hline
     &     $\varepsilon_{1s}$ [eV] ~~&~~   $\Gamma_{1s}$ [eV]\\
     \hline
     & & \\[-3mm]
% OR model  &  &  \\
% OR model  &  &  \\
      IA & $791$ & $2063$ \\
      FCA  & $1124$ & $626$ \\
     \hline
%         & \\[-3mm]
%BCN model     &  &  \\
%      IA & $406$ & $2013$\\
%      FCA & $1069$  &  $794$\\         
      \hline
%      \hline
    \end{tabular}
    \caption{Energy shift and width of the $1s$ atomic level in kaonic deuterium.}
    \label{tab:Kdatom}
\end{table}
We observe that the multi-scattering effects implemented in the FCA drastically modify the prediction for the $1s$ kaonic deuterium level, especially for its width, which gets reduced by more than a factor of three. Although we recognize that it is still too soon to compare with the SIDDHARTA2 measurement, our FCA result for the width seems to be in agreement with that of the preliminary analysis of \cite{SIDDHARTA-2_Sgaramella}, while our energy shift comes out about 30\% larger.

\subsection{\mbox{\boldmath $K^-d$} and \mbox{\boldmath $K^+d$} correlation functions}

\begin{figure*}
    \centering
\includegraphics[width=0.95\linewidth]{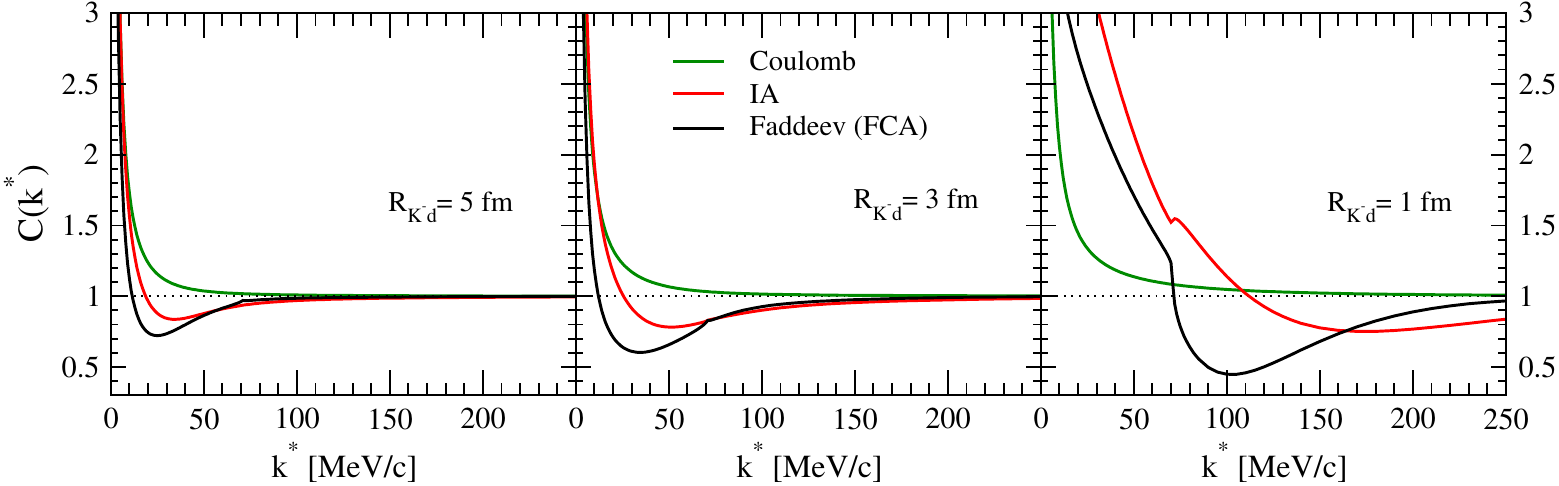}
    \caption{Correlation function $C(k^*)$ of $K^-d$ pairs for three different values of the radius $R_{Kd}$ of the Gaussian source. The green line accounts only for the Coulomb interaction, while the additional effects of the strong $K^-d$ amplitude are implemented within the IA (red line) or the FCA (black line).}
    \label{fig:corr_kmD_radii}
\end{figure*}

We start this section by presenting our results for the $K^- d$ correlation function. In Fig.~\ref{fig:corr_kmD_radii} we show $C(k^*)$ for three different Gaussian source sizes, where $k^*$ is the relative momentum in the c.m. of the pair. The Coulomb correlation function (green lines) is larger than one, reflecting the attractive character of the interaction. The correlation functions computed using the IA (red lines) and FCA (black lines) are notably different, emphasizing the importance of accounting for rescattering effects in this system. For a small radius (right panel), at low momenta, the correlation function appears above the Coulomb one, indicating that the localized source is selecting the low $r$ region of the wave function, which is very much enhanced by the existence of a bound ${\bar K}NN$ state close to the $K^-d$ threshold, a fact that is also reflected in the characteristic minimum observed for all source sizes~\footnote{See the illustrative discussion on this matter in Fig. 3 of Ref.~\cite{Liu:2023uly}}. As clearly seen in the results of Fig.~\ref{fig:corr_kmD_radii}, the correlation function captures the effect of the strong $K^- d$ interaction significantly,  but the sensitivity is gradually lost as the source radius increases.  In addition, one observes a subtle cusp structure around a c.m. momentum of $k^*=70$~MeV that is connected to the opening of the ${\bar K^0}nn$ channel on top of the already opened ${K^-}d$ one.    

\begin{figure*}
    \centering
\includegraphics[width=0.95\linewidth]{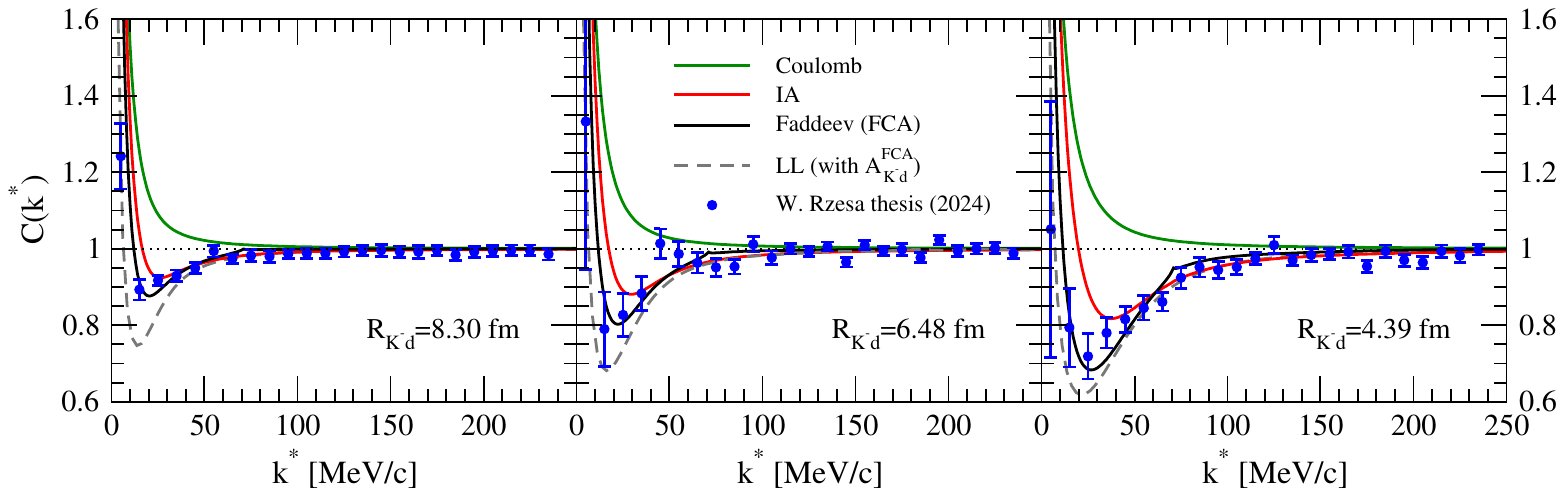}
    \caption{Correlation function $C(k^*)$ of $K^-d$ pairs compared with the
    experimental data taken from \cite{Rzesa:2024nra_thesis}.
    The green line accounts only for the Coulomb interaction, while the additional effects of the strong $K^-d$ amplitude are implemented within the IA (red line) or the FCA (black line).}
    \label{fig:corr_kmD}
\end{figure*}

Figure~\ref{fig:corr_kmD} complements the analysis by considering realistic source radii obtained from experimental data~\cite{Rzesa:2024nra_thesis}. This data was collected by the ALICE@LHC collaboration in Pb-Pb collisions at $\sqrt{s_{NN}}=5.02$ TeV for three different centralities: 0-10\% (central collisions), 10-30\% (mid-peripheral collisions), and 30-50 \% (peripheral collisions). The impact parameter of the collision (the distance between the centers of the initial nuclei) increases with the centrality class.
The comparison with measured correlation functions, including statistical uncertainties, demonstrates that the full consideration of rescattering effects encoded in the FCA achieves a significantly better agreement with experiment than the simple IA approach, particularly for smaller source sizes. These results highlight the relevance of multiple scattering processes within the deuteron. 

We note that, in the case of the correlation function for $R_{K^-d}=6.48$~fm,  our approach cannot reproduce the sharp rise around a momentum of 40 MeV, appearing right after the minimum that signals the presence of a subthreshold quasi-bound ${\bar K}NN$ state in the $T_{K^-d}$ amplitude. We do not find a reasonable physical interpretation for this rise in the experimental data, rather than associating it to a simple fluctuation, since this behavior is not seen for any of the two other multiplicities.

Apart from this exception, the general consistency between the theoretical predictions based on FCA and the experimental observations reinforces the reliability of this approach in capturing the essential dynamics of the $K^-d$ system.

%{\color{blue}
We finish the analysis of the $K^-d$ correlation function by commenting on the results shown by the dashed grey lines, obtained within the LL approximation and employing our FCA value of the scattering length, $A_{K^-d}=-2.06 +{\rm i} 1.77 $ fm. 
We observe that the LL correlation function, which only accounts for the asymptotic behavior of the wave function\footnote{The expression of the asymptotic wave function of a pair interacting via the strong and Coulomb forces can be found, for instance, in Eq.~(A2) of Ref.~\cite{Torres-Rincon:2024znb}}, differs from the full FCA result and does not reproduce the data satisfactorily. This example indicates that, in strongly correlated systems, analyzing the correlation function data within the LL approximation may be a too simplistic procedure.
%}

%\begin{figure*}
%    \centering
%\includegraphics[width=0.95\linewidth]
%{corr_kmD_BCN_Rmod_thesis.pdf}
%    \caption{BCN model}
%    \label{fig:corr_BCN_kmD}
%\end{figure*}

%\begin{figure*}
%    \centering
%\includegraphics[width=0.95\linewidth]
%{corr_kmD_OR_vs_BCN_thesis.pdf}
%    \caption{OR and BCN models}
%    \label{fig:corr_OR_BCN_kmD}
%\end{figure*}

%\begin{figure*}
%    \centering
%\includegraphics[width=0.95\linewidth]
%corr_kmD_OR_vs_BCN_thesis_2.pdf}
%    \caption{OR and BCN models, another option}
%    \label{fig:corr_OR_BCN_kmD_2}
%\end{figure*}

\begin{figure*}
    \centering
 \includegraphics[width=0.95\linewidth]{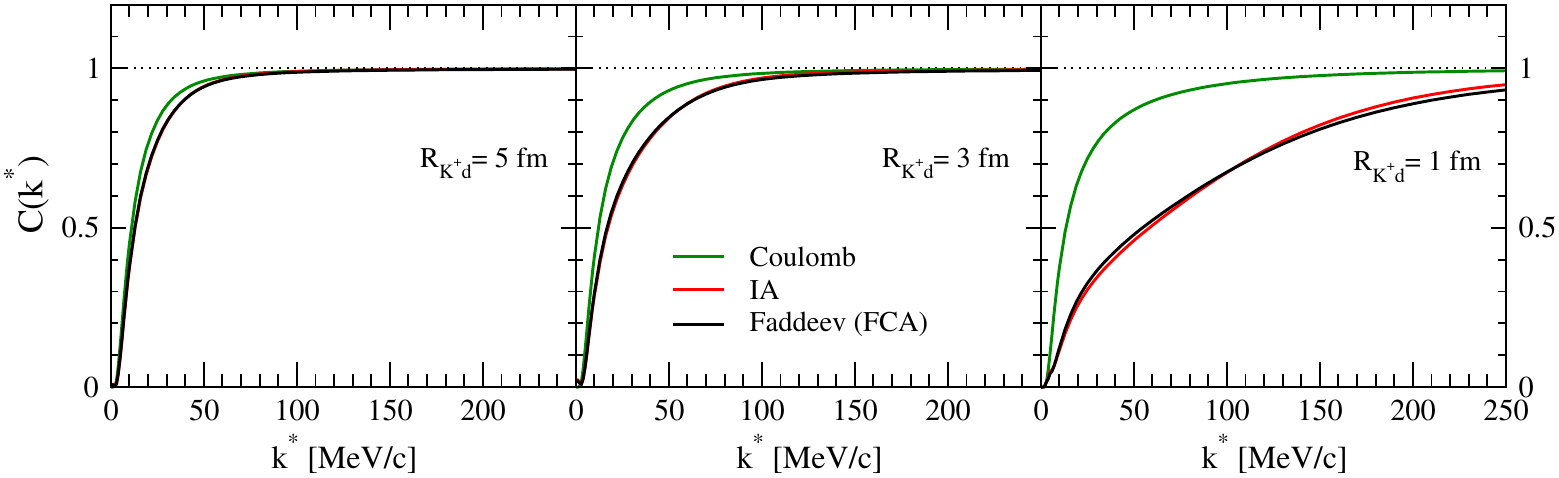}
    \caption{Correlation function $C(k^*)$ of $K^+d$ pairs for three different values of the radius $R_{Kd}$ of the Gaussian source. The green line accounts only for the Coulomb interaction, while the additional effects of the strong $K^+d$ amplitude are implemented within the IA (red line) or the FCA (black line).}
    \label{fig:corr_kpD_radii}
\end{figure*}

%\begin{figure}
%    \centering
%\includegraphics[width=0.95\linewidth]{corr_kmD_r135.pdf}
 %   \caption{OR and BCN models, r=1.35 fm}
 %   \label{fig:corr_kmD_r135}
%\end{figure}
We now turn to the discussion of the correlation function of $K^+ d$ pairs and again we show our results for three different source sizes in Fig.~\ref{fig:corr_kpD_radii}. This time, the Coulomb-only correlation function (green line) is smaller than one, as corresponds to a repulsive interaction. There is no substantial difference between the results including the strong effects within the IA (red lines) or the FCA (black lines), indicating that rescattering effects are not important in this case because of the weaker $KN$ interaction. The effect of the repulsive $K^+ d$ interaction is better seen for small values of the source radius, which captures the low distance behavior of the wave function, where it shows more differences with respect to the Coulomb-only one. The correlation function rapidly loses sensitivity to the strong interaction as the radius increases, and it practically coincides with the Coulomb one for a source radius of 5~fm.

\begin{figure}
    \centering
\includegraphics[width=0.8\linewidth]{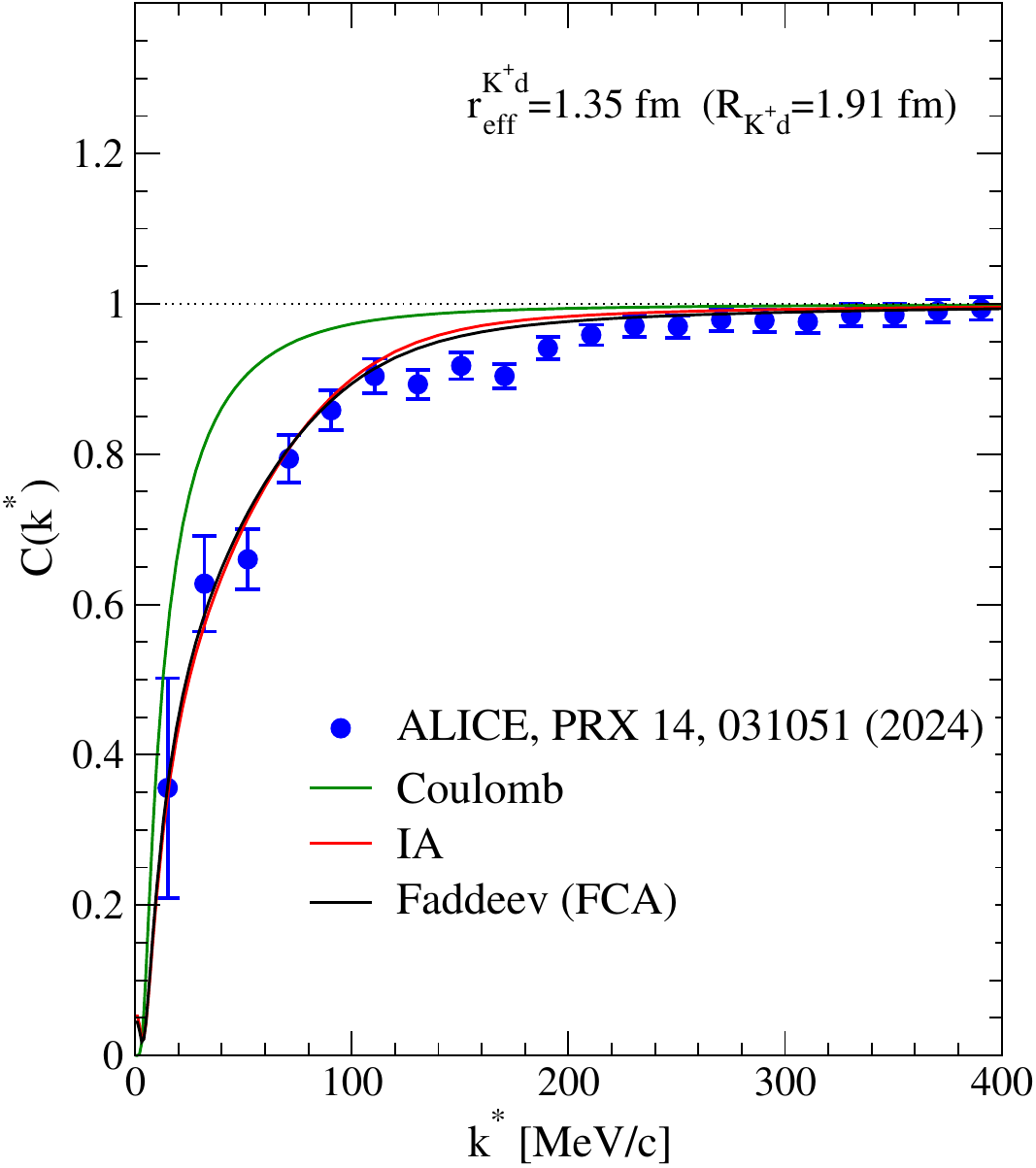}
    \caption{
    Correlation function $C(k^*)$ of $K^+d$ pairs compared with the
    experimental data taken from \cite{ALICE:2023bny}.
    The green line accounts only for the Coulomb interaction, while the additional effects of the strong $K^-d$ amplitude are implemented within the IA (red line) or the FCA (black line).
    }
    \label{fig:corr_kpd_r135}
\end{figure}

 In Fig.~\ref{fig:corr_kpd_r135}, our results are compared to the ALICE experimental data~\cite{ALICE:2023bny} measured in high-multiplicity p-p collisions at $\sqrt{s}=13$ TeV. While there is no well-defined definition of centrality in proton-proton collisions, the relevant events are chosen among those with the highest multiplicity of final charged particles. Our results show excellent agreement with the data, confirming the validity of our strong interaction model, which has a weak repulsive character.
 
\begin{figure*}
    \centering
 \includegraphics[width=0.95\linewidth]{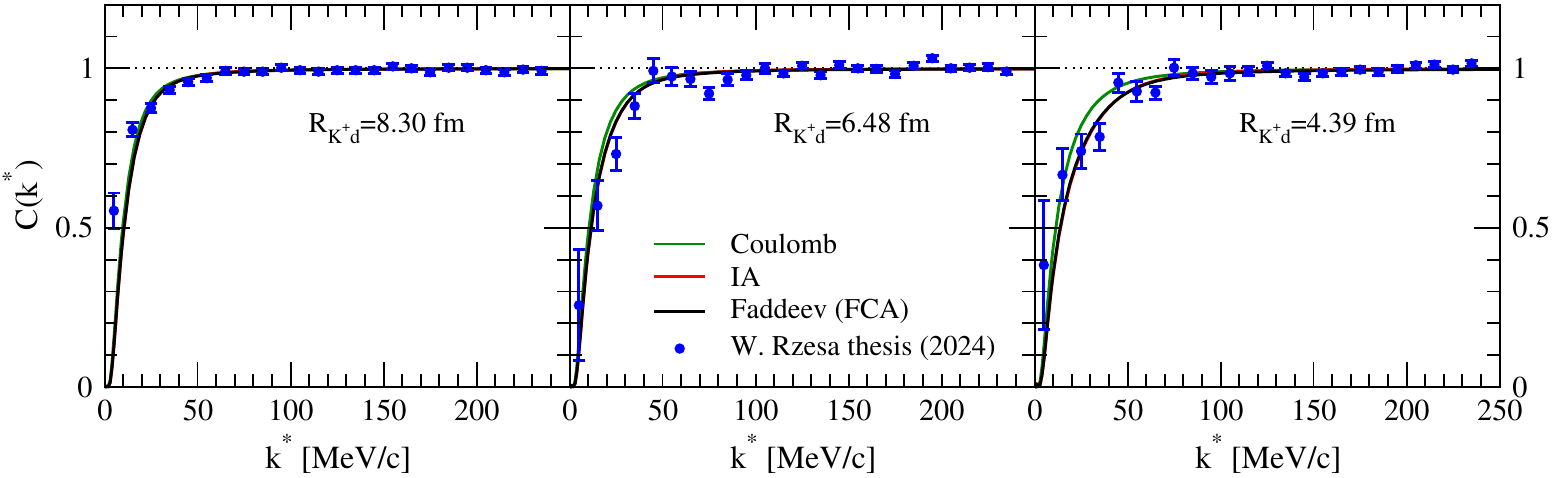}
    \caption{Correlation function $C(k^*)$ of $K^+d$ pairs compared with the
    experimental data in p-p collisions taken from~\cite{Rzesa:2024nra_thesis}. The green line accounts only for the Coulomb interaction, while the additional effects of the strong $K^+d$ amplitude are implemented within the IA (red line) or the FCA (black line).}
    \label{fig:corr_kpD}
\end{figure*}

Finally, in Fig.~\ref{fig:corr_kpD}, we compare our correlation functions with experimental data obtained considering the source radii obtained in the analysis presented in Rzesa's PhD thesis ~\cite{Rzesa:2024nra_thesis}. This data was collected in the ALICE@LHC collaboration in Pb-Pb collisions at $\sqrt{s_{NN}}=5.02$ TeV, the same system as in Fig.~\ref{fig:corr_kmD} but this time for $K^+d$ pairs. The agreement with data is very good but the uncertainties do not allow to distinguish between the results with or without the strong interaction effects for these large values of the source size. We hope that future measurements with improved statistical precision will help clarify the role of the strong interaction in the $K^+d$ system and provide further constraints for theoretical models.

\section{Conclusions}
\label{sec:conclusions}

In this work, we have developed a unified theoretical framework to investigate kaon-deuteron ($K^-d$ and $K^+d$) femtoscopy by combining effective chiral interactions for the two-body $K^-N$ and $K^+N$ sectors and the Faddeev equations in the impulse approximation and the fixed-center approximation. Our model incorporates both the strong and Coulomb interactions and evaluates the correlation functions using the Koonin-Pratt formalism.

Our results show that the $K^- d$ correlation function is highly sensitive to the size of the source, reflecting the strong attractive interaction dominated by a subthreshold ${\bar K}NN$ resonance ($\sqrt{s} \simeq 2360$ MeV), which is originated from the $\Lambda (1405)$ in combination with an additional nucleon. We observe the importance of multi-scattering effects in this sector, evidenced by the significant deviation from the Coulomb baseline, an enhanced minimum of the correlation and a cusp structure. In contrast, the $K^+d$ system is governed by a weakly repulsive and predominantly elastic interaction, presenting small differences between the impulse and the fixed-center approximations. Deviations with respect to the Coulomb result are only evident at small source radii.

Our theoretical correlation functions for both $K^-d$ and $K^+d$ pairs were found to reproduce the preliminary experimental measurements of the ALICE collaboration in various collision systems. 
For smaller radii, the present level of precision of the $K^-d$ femtoscopy data is already able to discard the results of the simpler impulse approximation in favor of those that include the multiple-scattering processes.
This reinforces the predictive power of our approach and underscores the role of femtoscopy as a valuable probe of hadronic interactions involving strangeness.

Furthermore, predictions for the energy shift and width of kaonic deuterium also highlight the importance of multiple scattering in shaping atomic-level observables, with our results aligning with preliminary indications from the SIDDHARTA2 experiment. 

In conclusion, the global analysis of scattering amplitudes, femtoscopy correlation functions, and kaonic atom properties demonstrates the necessity of incorporating coupled-channel chiral dynamics and rescattering effects in modeling the $K^-d$ and $K^+d$ interactions.
While we have captured the fundamental physics and the relevant dynamics in our calculation, we will refine and test the various approximations we have made in this work, e.g. baryon static approximation, neglecting the breakup processes or the factorization of the strong and Coulomb scattering amplitudes. These studies will be addressed in future works.

%In conclusion, this work validates the chiral effective theory combined with the fixed-center approximation as a robust tool for describing $K^-d$ and $K^+d$ interactions. While we have captured the fundamental physics and the relevant dynamics in our calculation, we will refine and test the various approximations we have made in this work, e.g. baryon static approximation, neglecting the breakup processes or the factorization of the strong and Coulomb scattering amplitudes. These studies will be addressed in future works. 

\begin{acknowledgments}

This work has been supported by the project numbers CEX2024-001451-M (Unidad de Excelencia ``Mar\'ia de Maeztu'') and PID2023-147112NB-C21, funded by the Spanish MCIN/ AEI/10.13039/501100011033/; and by Contract 2021 SGR 171 by the Generalitat de Catalunya.
JMT-R also thanks the CRC-TR 211 ’Strong-interaction matter under extreme conditions’, Project No. 315477589 - TRR 211 by the Deutsche Forschungsgemeinshaft, and Grant No. 402942/2024-8 by the Brazilian CNPq (National Council for Scientific and Technological
Development).
\end{acknowledgments}

\bibliography{references}

\end{document}